\documentclass[usenatbib]{mn2e}
\usepackage{graphicx}
\usepackage{float}

\def\go{
\mathrel{\raise.3ex\hbox{$>$}\mkern-14mu\lower0.6ex\hbox{$\sim$}}
}
\def\lo{
\mathrel{\raise.3ex\hbox{$<$}\mkern-14mu\lower0.6ex\hbox{$\sim$}}
}
\def\simeq{
\mathrel{\raise.3ex\hbox{$\sim$}\mkern-14mu\lower0.4ex\hbox{$-$}}
}

\def\etal{{et al.\ }}

\begin{document}

\title[The unusual X-ray light-curve of GRB 080307]
{The unusual X-ray light-curve of GRB 080307: the onset of the afterglow?}
\author[K.L. Page \etal]{K.L. Page$^{1}$, R. Willingale$^{1}$, P.T. O'Brien$^{1}$, N.R. Tanvir$^{1}$, J.P. Osborne$^{1}$, B. Zhang$^{2}$, \and S.T. Holland$^{3,4}$, A.J. Levan$^{5}$, A. Melandri$^{6}$, R.L.C. Starling$^{1}$, D. Bersier$^{6}$, \and D.N. Burrows$^{7}$, J.E. Geach$^{8}$ \& P. Maxted$^{9}$\\
$^{1}$ X-Ray and Observational Astronomy Group, Department of Physics \&
  Astronomy, University of Leicester, Leicester, LE1 7RH, UK\\
$^{2}$ Department of Physics, University of Nevada, Las Vegas, Nevada, USA\\
$^{3}$ NASA Goddard Space Flight Center, Greenbelt, MD 20771, USA\\
$^{4}$ Universities Space Research Association\\
$^{5}$ Department of Physics, University of Warwick, Coventry, CV4 7AL, UK\\
$^{6}$ Astrophysics Research Institute, Liverpool John Moores University, Twelve Quays House, Birkenhead, CH41 1LD, UK\\
$^{7}$ Department of Astronomy and Astrophysics, Pennsylvania State University, State College, PA 16802, USA\\
$^{8}$ Department of Physics, Durham University, Durham, DH1 3LE, UK\\
$^{9}$ Astrophysics Group, Keele University, Keele, ST5 5BG, UK
}

\label{firstpage}

\maketitle

\begin{abstract}

{\it Swift}-detected GRB~080307 showed an unusual smooth rise in its X-ray light-curve around 100 seconds after the burst, at the start of which the emission briefly softened. This `hump' has a longer duration than is normal for a flare at early times and does not demonstrate a typical flare profile. Using a two component power-law-to-exponential model, the rising emission can be modelled as the onset of the afterglow, something which is very rarely seen in {\it Swift}-X-ray light-curves. We cannot, however, rule out that the hump is a particularly slow early-time flare, or that it is caused by upscattered reverse shock electrons.

\end{abstract}


\date{Received / Accepted}

\section{Introduction}
\label{intro}

Although many Gamma-Ray Burst (GRB) X-ray light-curves roughly follow a `canonical' decay of initially steep and then
flat (the so-called `plateau' phase), before settling into a classical power-law decline phase (see, e.g. Nousek et al. 2006; Zhang et al. 2006), some are noticeably
different. Examples include GRB~060105, which had an unusually shallow initial decay (Godet et al. 2008);
GRB~060218, which was clearly associated with supernova 2006aj and showed the
break-out of a shock wave from the exploding progenitor (Campana et al. 2006);
GRB~061007, where the X-ray decay continued with no
breaks until at least 10$^{6}$~seconds after the trigger (Schady et al. 2007; Mundell et al. 2007);
GRB~070110, which showed an abrupt drop in the X-ray emission at the end of
the `plateau' stage (Troja et al. 2008). It is examples such as these, whose behaviour differs from those of more typical afterglows, which provide further insight into the physics of the bursts, allowing the testing of models for the early stages of GRB afterglows.

GRB~080307 has a very unusual X-ray light-curve, reminiscent of GRB~060218. The emission rose
smoothly until about 200~seconds after the trigger, at which point a long, steady
decay set in, continuing unbroken until the afterglow could no longer be detected after a few 10$^{5}$~s. Here we show that, in contrast to GRB~060218, the early emission peak is probably the rising afterglow emission; there is no evidence for a supernova shock breakout.

The observations are described in Section~\ref{obs}, with details on
the {\it Swift} spectral analysis and results in Sections~\ref{bat}--\ref{uvot}; ground-based optical follow-up findings are given in
Section~\ref{ground}. Section~\ref{disc} presents a discussion.

\section{Observations}
\label{obs}

\subsection{Swift}

Following the BAT trigger at 11:23:30 UT on 2008-03-07, {\it Swift} (Gehrels et
al. 2004) immediately
slewed to the burst and started collecting X-ray Telescope (XRT; Burrows et
al. 2005) and UV/Optical Telescope (UVOT; Roming et al. 2005) data at T+99~s and
T+105~s respectively. The XRT centroided onboard, though a refined position
was quickly determined from the promptly-downlinked SPER (Single Pixel Event
Report) data (Holland et al. 2008). The most accurate {\it Swift} position was later
calculated by using the XRT-UVOT alignment and matching UVOT field sources
to the USNO-B1 catalogue (see Goad et al. 2007b for more details): R.A.~=~09$^h$06$^m$30.72$^s$,
Dec.~=~+35$^o$08'20.3" (J2000; 90~per~cent confidence radius of 1.8 arcsec;
Goad et al. 2008).

The data were processed with the standard {\it Swift} FTools, using v2.8 of
the software (release date 2007-12-06) and the corresponding calibration files. The XRT stayed
in Windowed Timing (WT) mode until around 614~s after the trigger, at which
point the count-rate dropped below 2~count~s$^{-1}$ and the XRT automatically
switched to Photon
Counting (PC) mode. The PC data obtained towards the end of the first orbit were
piled-up, so the light-curve and spectrum were extracted using an annulus to
exclude the central four pixels of the Point Spread Function [PSF; the exclusion
radius was estimated by fitting the wings of the PSF with the known King
function parameters (Moretti et al. 2005) and determining at which point the observed profile deviated from the
expected model]. A comparison of ancillary response files with and without the
PSF correction provided the factor by which the count rate needed to be scaled
to account for the counts excluded by the annulus.

\subsection{Ground-based follow-up}

Optical observations were made with GMOS (Gemini Multi-Object Spectrograph) on Gemini North (Tanvir 2008), the Auxilliary Port Camera
on the William Herschel Telescope (WHT/AUX) and the 2-m Faulkes Telescope South (FTS). UKIRT/WFCAM (UK Infrared Telescope/Wide Field Camera) also observed and detected the afterglow in the IR; in this case, a sequence of exposures was obtained, cycling through the filters $JHJKJHJ$, until the target hit the altitude limit of the telescope. 


The optical and near-IR magnitudes measured are provided in Table~\ref{mag}; these have not been corrected for the reddening of E$_{\rm B-V}$~=~0.03 mag (Schlegel, Finkbeiner \& Davis 1998).
 UKIRT magnitudes were calibrated against 15 2MASS stars in the field. The WHT and FTS-$i'$ data were calibrated using the SDSS system (Cool et al. 2008), while the USNOB1-0 catalogue was used for the $B$ and $R$ Bessell filters (Monet et al. 2003).

\begin{figure}
\begin{center}
\includegraphics[clip, width=8cm]{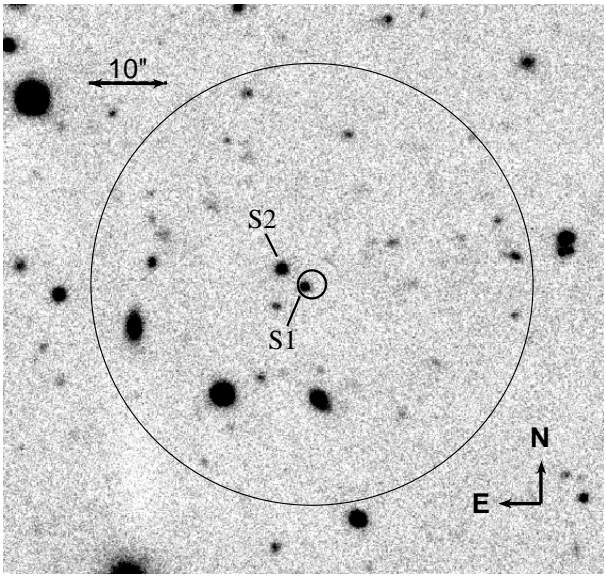}
\caption{Gemini image, showing the position of the optical afterglow (S1) and the nearby source which was found to contaminate the {\it Swift}-XRT data (S2). The large circle is the XRT extraction region (radius 28 arcsec).}
\label{fov}
\end{center}
\end{figure}

The Gemini data revealed a faint point source
within the revised XRT error circle (Goad et al. 2008), providing an accurate position of R.A.~=~09$^h$06$^m$30.80$^s$,
Dec.~=~+35$^o$08'20.1" (J2000; uncertainty of 0.2 arcsec) for the optical afterglow. Figure~\ref{fov} shows the Gemini image (S1 is the position of the afterglow; S2 is the contaminating source discussed below).










\begin{table}
\begin{center}
\caption{Optical and nIR measurements. No correction for the expected extinction corresponding to a reddening of E$_{\rm B-V}$~=~0.03 mag has been made.} 
\label{mag}
\begin{tabular}{p{1truecm}p{2.5truecm}p{1.5truecm}p{2.0truecm}}
\hline
Filter  & Magnitude& Time  & Telescope\\ 
 & & (min since burst)\\
\hline

$B$ & $>$21.3 (5 $\sigma$) & 36.2 & FTS \\
$r$ &  24.05 $\pm$ 0.22 & 630 & WHT\\
  &  24.43 $ \pm$ 0.11 & 36560 & WHT\\
 &  24.10 $\pm$ 0.25 & 76896 & WHT\\
$R$  & 21.60~$\pm$~0.25 &  22.2 & FTS\\
 &  22.02~$\pm$~0.47 & 52.1 & FTS\\
$i$ & 22.28 $\pm$ 0.03 & 89 & Gemini\\
  &      22.32 $\pm$ 0.03&    93 & Gemini \\
  &      22.38 $\pm$ 0.03&    97 & Gemini \\
  &      22.36 $\pm$ 0.03&    101& Gemini \\
  &      22.38 $\pm$ 0.03&    105& Gemini \\
  &      23.66 $\pm$ 0.19 & 640 & WHT\\
  &      24.34 $\pm$ 0.26 & 36640 & WHT\\
$i'$ & 20.46~$\pm$~0.17 & 34.78 & FTS\\
\hline
 $J$ & 19.74 $\pm$ 0.10 & 44 & UKIRT\\
 &19.96 $\pm$ 0.11 & 54 & UKIRT\\
  &20.74 $\pm$ 0.22 & 64 & UKIRT\\
   &20.24 $\pm$ 0.29 & 73 & UKIRT\\
 $H$  & 19.41 $\pm$ 0.12 & 49 & UKIRT\\
 & 19.29 $\pm$ 0.13 & 69 & UKIRT\\
$K$ & 18.14 $\pm$ 0.08 & 59 & UKIRT\\

\hline

\end{tabular}
\end{center}

\end{table}

\subsection{Chandra}
\label{contam}

During an initial inspection of the XRT data, the X-ray light-curve showed a break at around 4~$\times$~10$^{4}$~s after the trigger. An apparent very flat decay then continued until the end of the {\it Swift} observations, around 6~$\times$~10$^{6}$~s (Holland et al. 2008b). Because Gemini optical imaging showed the presence of a nearby object (Figure~\ref{fov}), just a few arcseconds from the GRB position, we obtained a late-time {\it Chandra} Target of Opportunity observation  in September 2008, 6.5 months after the burst, to determine whether the slow X-ray decline was intrinsic or due to a contaminating source.

The {\it Chandra} data did reveal that there was an X-ray source
coincident with the optical one, well within the point-spread function of the XRT. This contaminating source is at a position of R.A.~=~09$^{h}$06$^{m}$31.04$^{s}$, Dec.~=~+35$^{o}$08'22.4'' (J2000), 3.8~arcsec from the afterglow position (determined from the optical data). 
Its $i$-band magnitude is $\sim$~21.85 [$J$~=~20.67, $H$~=~19.88, $K$~=~18.93 from UKIRT observations, with uncertainties of $\sim$0.16 mag], and the X-ray flux is approximately 2.7~$\times$~10$^{-14}$ erg~cm$^{-2}$~s$^{-1}$ (observed; 3.8~$\times$~10$^{-14}$ erg~cm$^{-2}$~s$^{-1}$ unabsorbed). This implies an optical-to-X-ray ratio of $\alpha_{\rm ox}$~$\sim$~1.4, typical for an active galactic nucleus. The afterglow itself was not detected during the {\it Chandra} pointing.

The flux of the nearby source corresponds to $\sim$~8~$\times$~10$^{-4}$~count~s$^{-1}$ in the XRT. This value has been subtracted off all the X-ray data points in the light-curve. 


At the flux level of the contaminating source, the sky density of X-ray sources is about 57 per square degree (Mateos et al. 2008). This gives approximately a 1\% chance of finding such a source in the extraction region used (radius of 28 arcsec). Given that {\it Swift} has detected more than 350 GRBs to date, it is likely that other X-ray light-curves reaching similar levels have been affected by contaminating sources. A possible example, noted by the authors, is GRB~050422 (Beardmore et al. 2007), which also showed a levelling-off of the light-curve at later times.

\section{Analysis and Results}
\label{anal}

\subsection{Swift-BAT}
\label{bat}

The Burst Alert Telescope (BAT; Barthelmy et al. 2005) data give a T$_{90}$ value of 126~$\pm$~26 s. A spectrum
covering this time can be fitted with
a single power-law photon index of $\Gamma$~=~1.81$^{+0.24}_{-0.22}$, with no requirement
for any form of energy cut-off or break. The corresponding 15--150~keV flux is
6.3~$\times$~10$^{-9}$ erg~cm$^{-2}$~s$^{-1}$. 

\begin{figure}
\begin{center}
\includegraphics[clip, width=8cm]{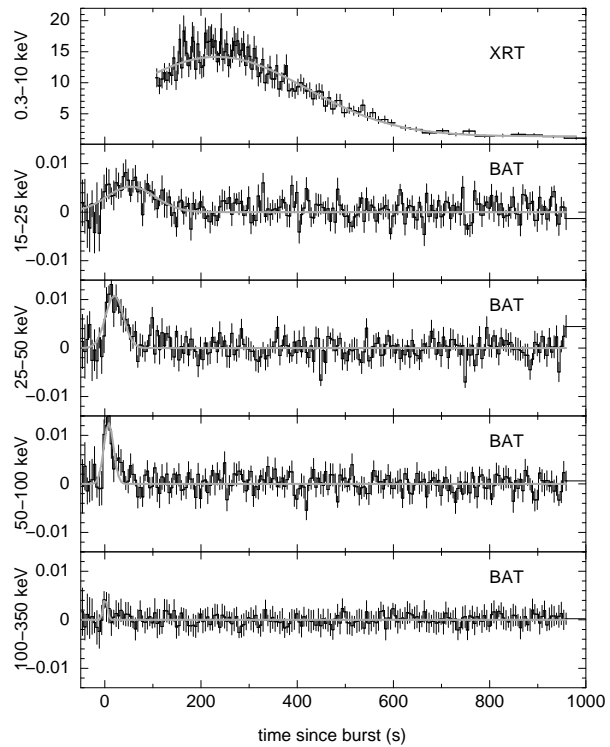}
\caption{The light-curve peak becomes broader and moves later in time over the softer bands - see
  Table~\ref{peak}. The model fitted is a Gaussian (together with an underlying power-law decay for the XRT data). The units are count~s$^{-1}$ (fully illuminated detector)$^{-1}$ for the BAT and count~s$^{-1}$ for the XRT.}
\label{alllc}
\end{center}
\end{figure}

\begin{figure}
\begin{center}
\includegraphics[clip, angle=-90, width=8cm]{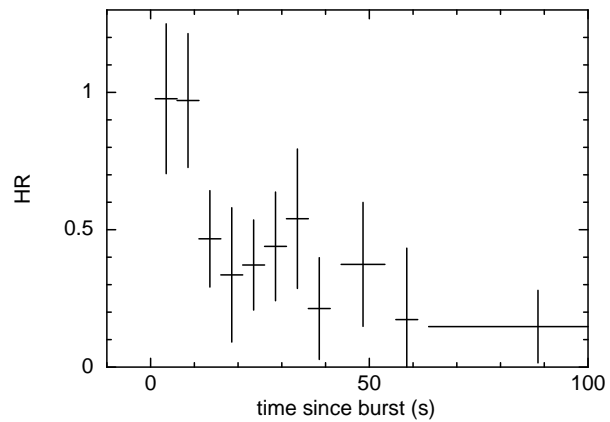}
\caption{The BAT hardness ratio (HR) of GRB~080307, showing a spectral softening over the first 100~s of the emission. The HR is defined as the ratio of the counts in the 50--150~keV and 15--50~keV bands.}
\label{bathr}
\end{center}
\end{figure}

Figure~\ref{alllc} shows the BAT light-curves over the standard energy bins (15--25, 25--50, 50--100 and 100--350~keV), while
Figure~\ref{bathr} demonstrates
how the $\gamma$-ray emission softened during the burst. Fitting spectra for
T~=~0--50~s and 50--126~s also shows this trend:
$\Gamma_{\rm 0-50s}$~=~1.53~$\pm$~0.18 and $\Gamma_{\rm 50-125s}$~=~3.18$^{+0.96}_{-0.70}$.

Figure~\ref{alllc} also shows how the peak
of the $\gamma$-ray data both moved to later times and broadened at softer
energies; this trend continued for the X-ray data. A Gaussian component was applied to each light-curve to parametrize the changes in a simple manner and the results are given in Table~\ref{peak}.

\begin{table}
\begin{center}
\caption{Gaussian component fitted to each energy band of the {\it Swift} light-curve (Figure~\ref{alllc}). The peak is narrower and occurs earlier at harder energies.} 
\label{peak}
\begin{tabular}{p{3.0truecm}p{2.0truecm}p{1.5truecm}}
\hline
Bandpass (keV) & Gaussian  & Width (s)\\ 
& centre (s)\\
\hline
0.3--10 & 234$^{+16}_{-17}$ & 185~$\pm$~15\\  
15--25 & 54~$\pm$~9 & 39$^{+10}_{-8}$\\
25--50 & 23$^{+5}_{-4}$ & 17$^{+5}_{-4}$\\
50--100 & 8.6$^{+1.5}_{-1.3}$ & 6.6$^{+1.8}_{-1.3}$\\
100-350 & 1.3$^{+2.4}_{-1.0}$ & 1.5$^{+2.6}_{-0.9}$\\

\hline
\end{tabular}
\end{center}
\end{table}

\subsection{X-ray analysis}
\label{xrt}

\subsubsection{Temporal Analysis}
\label{temp}


Excluding the hump in the light-curve and fitting the XRT data after 800~s, a single power-law is a good fit, with a decay slope of $\alpha$~=~1.95~$\pm$~0.10
(where F$_{\nu,t}$~$\propto$~$\nu^{-\beta}$t$^{-\alpha}$ and the photon spectral index, $\Gamma$~=~$\beta$+1). After about 2~$\times$~10$^{5}$~s, the afterglow is no longer significantly detected above the nearby source. A slope of $\alpha$~$\sim$~1.95 is more typical of the `normal' decay phase of the `canonical' X-ray afterglow (Nousek et al. 2006; Zhang et al. 2006; Evans et al. 2009; Racusin et al. 2009), being significantly steeper than the `plateau' stage.


The fluence of the hump emission between
100--600~s is $\sim$~2.1~$\times$~10$^{-7}$
erg~cm$^{-2}$, about a quarter of the time-averaged BAT (15--150~keV) burst fluence
measured by Sato et al. (2008).

\begin{figure}
\begin{center}
\includegraphics[clip, width=8cm]{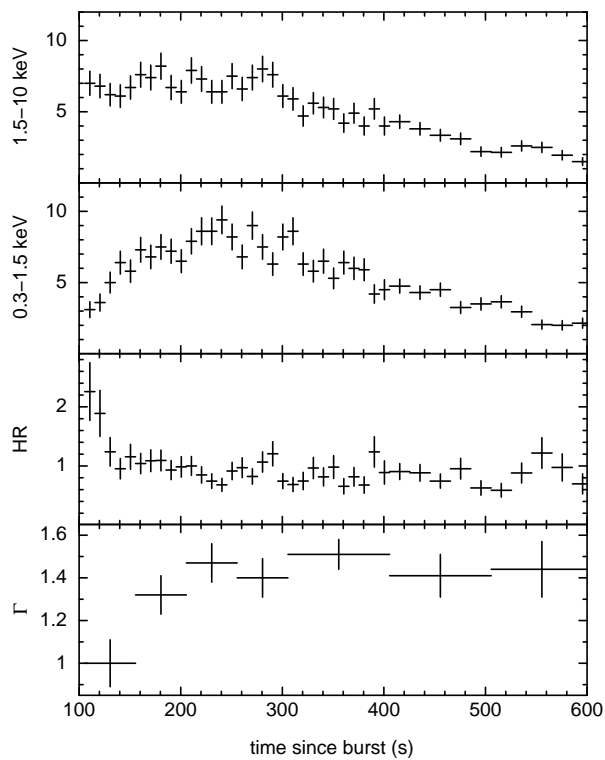}
\caption{The XRT-WT light-curves over 1.5--10 and 0.3--1.5~keV, together with the hardness ratio, showing an initial softening until
  $\sim$~140~s. After this time the hardness ratio remains approximately constant. The bottom panel shows the photon indices obtained from fitting the spectra with a single power-law.}
\label{xrthr}
\end{center}
\end{figure}

Figure~\ref{xrthr} compares the hard (1.5--10~keV) and soft (0.3--1.5~keV)
light-curves, showing there is a softening during the beginning of the
brightening phase, after which the hardness ratio remains close to constant;
the data beyond 600~s show no evidence for a change in hardness either. This early
behaviour is
unusual for a GRB `flare' or `pulse': typically, as the count-rate increases, so, too, does the hardness
(e.g., Golenetskii et al. 1983; Ford et al. 1995; Borgonovo \& Ryde 2001; Goad
et al. 2007a; Page et al. 2007). However, the start of this rise may not have been caught by the XRT, so there could have been an earlier hardening -- and some flares do seem to start hard and then just soften (e.g., GRB~050502B, Falcone et al. 2006). The spectral evolution of flares does tend to continue during the peak and beyond, though, which is not seen in this case; after about 140~s post-trigger, there is little or no further change of spectral shape in GRB~080307.


\subsubsection{Spectral Analysis}
\label{spec}

Time-sliced spectra were extracted throughout the hump in the light-curve from T$_0$+100 -- 600~s, and the resulting power-law fits are shown in the bottom panel of Figure~\ref{xrthr}. The absorbing column was fixed at the Galactic value 
(2.37~$\times$~10$^{20}$~cm$^{-2}$; Kalberla et al. 2005), since no higher N$_{\rm H}$ was suggested by the fit. An alternative fit of keeping the power-law index fixed and allowing the variation of N$_{\rm H}$ to account for the difference in spectral shape did not provide such a good fit ($\chi^2$ was 10 higher, for one fewer degree of freedom).

A spectrum extracted for the PC data from T$_0$+4.4~ks onwards (to avoid any residual emission from the `flare') has $\Gamma$~=~1.66~$\pm$~0.30. Note, however, that the fit is improved if N$_{\rm H}$ is allowed to vary (99.5\% via the F-test), leading to $\Gamma$~=~2.16$^{+0.36}_{-0.33}$, with a total column at z~=~0 (including the Galactic value) of N$_{\rm H}$~=~(1.8$^{+2.2}_{-1.5}$)~$\times$~10$^{21}$~cm$^{-2}$. This is discussed further in Section~\ref{disc}.




\subsection{Swift-UVOT}
\label{uvot}

No afterglow was detected by the UV/Optical Telescope (UVOT; Roming et
al. 2005). Holland (2008) lists the 3$\sigma$ upper limits, which range between
$>$~20.6--21.4 for the optical and UV filters (summed over a few hundred to a few thousand seconds after the trigger). The afterglow was also not detected in the
broad-bandpass white filter, to a limiting magnitude of $>$22.3.

\subsection{Ground-based follow-up}
\label{ground}

Gemini detected a faint $i\approx22.3$ point source approximately 90
minutes post-trigger (Figure~\ref{fov}), which faded by about 0.1 mag during the course
of this sequence (Table~\ref{mag}). WHT also found a faint blue source approximately 11 hours post-burst in $r$ and $i$ filters, although subsequent observations with the WHT over the course of
several weeks showed no significant variation in the luminosity
of the source, indicating the host galaxy was being detected.

\begin{figure}
\begin{center}
\includegraphics[clip, angle=-90, width=8cm]{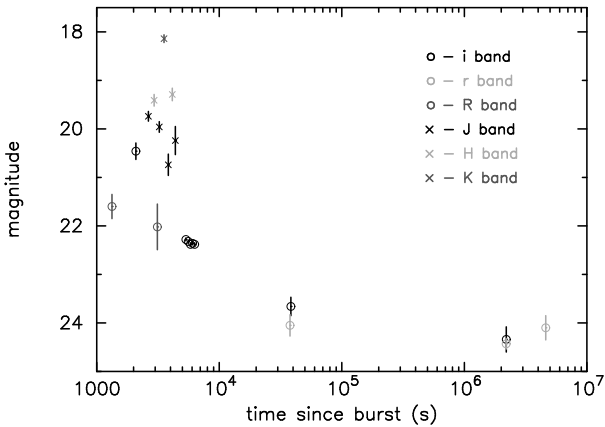}
\caption{Ground-based optical data obtained using Gemini/GMOS ($i$-band), WHT/AUX ($i$ and $r$-band) and FTS ($R$ and $i$) and near IR ($J$, $H$ and $K$-band) data from UKIRT.}
\label{optlc}
\end{center}
\end{figure}

Figure~\ref{optlc} plots the optical and IR data listed in Table~\ref{mag}.
The second UKIRT $H$-band measurement seems to be slightly in excess of the previous point, and the $J$-band point around 4.2~ks is above that at $\sim$~3.8~ks. Unfortunately there are no simultaneous X-ray observations, but the X-ray data at the beginning of the second orbit ($\sim$~4.4~ks) do seem to be slightly higher than the underlying power-law. It is therefore plausible that there was a similar fluctuation in the X-ray and IR bands around this time.

\section{Discussion}
\label{disc}

\subsection{Limits on a supernova component}

The late time limit on variability between the final two r-band
observations is 0.23~$\pm$~0.22~$\mu$Jy, corresponding to a 3$\sigma$
limit on any variable source of 0.66 $\mu$Jy, or an $r$~$\sim$~24.4
source, which we take to be the limit of any supernova present at the
time of our intermediate r-band observations (roughly 25 days
postburst in the observer frame). Comparing this to redshifted
lightcurves of SN~1998bw (Galama et al. 1998), we determine that we
would have observed a brightening due to a supernova component akin to
SN~1998bw should it have been located at $z \lo 0.7$. This could be
reduced if there were to be significant foreground extinction in the
direction of GRB 080307. The moderately red R-K colour (R-K $\sim$ 4),
suggests this may be the case, although there is no strong evidence for
excess $N_{\rm H}$ from the X-ray data (see Section~\ref{specanom}); thus,
the lack of SN component does not place strong constraints on the
redshift.

\subsection{Possible constraints on the redshift}

The lack of variation in the WHT data 
implies that those
observations are dominated by the host galaxy.
The absence of a supernova makes it unlikely that this burst occurred at low redshift [very high foreground extinction ($A_V$~$\sim$~4--5 for a burst at $z~=~0.1$) would be required], despite the resemblance between this light-curve and that of the low-$z$ 
GRB 060218 (Campana et al. 2006), whilst the magnitude of
the host galaxy is entirely consistent with many other GRB
hosts at $z\la2$ (see, e.g., Savaglio, Glazebrook \& Le Borgne 2008).

Ukwatta et al. (2008) have investigated certain BAT parameters which may
indicate when a burst is at a high redshift. Based on a limited sample of bursts, there are four criteria which together imply an 85\% chance of a GRB having a redshift greater than 3.5; GRB~080307 fulfils these four points. Note, though,  that the most distant GRB yet measured (GRB~080913; Greiner et al. 2008) did not satisfy all these criteria.

The XRT data from the second orbit onwards, when no spectral evolution is apparent and the light-curve is following a steady decline, gave an indication of an absorbing column in excess of the Galactic value of N$_{\rm H}$~$\sim$~1.6~$\times$~10$^{21}$~cm$^{-2}$ (at $z$~=~0); this is quite poorly constrained, however. This excess implies an upper limit on the redshift of $z$~$\la$~3.8 according to the method of Grupe et al. (2007a), though this increases to $z$~$\la$~6.5 using the lower limit on the excess column. This is in comparison to the lower limit estimate of $z$~$\ga$3.5 as indicated by the BAT data (Section~\ref{bat}), although the detection of the host galaxy makes it unlikely that the burst occurred at such a redshift.

\subsection{Spectral anomalies}
\label{specanom}

If the additional column of  N$_{\rm H}$~$\sim$~1.6~$\times$~10$^{21}$~cm$^{-2}$ is used when fitting time-sliced data through the hump, it is found that a single power-law is a poor fit, with a much softer component (either a second power-law, or a blackbody) being required as well. The `underlying' power-law still shows the softening trend plotted in Figure~\ref{xrthr}, though.

GRB~060218 showed a similar rise in its early X-ray light-curve, and it was found that the spectra during this interval showed evidence for an expanding thermal component, corresponding to a supernova shock breakout (Campana et al. 2006). Fitting a similar model here, however, results in the temperature and radius of the blackbody component
remaining approximately constant throughout (kT~$\sim$~130--140~eV and an emitting radius of $\sim$~200D$_{\rm 10 kpc}$~km,
where D$_{\rm 10 kpc}$ is the distance from the observer to the burst in units of 10~kpc), making it inconsistent with a shock breakout (a cooling temperature and expanding radius would be expected). No supernova has been identified for GRB~080307 either. It therefore seems unlikely that a thermal component is the explanation for the possible spectral curvature.  A dual power-law model can equally well fit the spectrum, with photon indices of $\Gamma$~$\sim$~3 and $\sim$~1 (both initially softening). Moretti et al. (2008) investigated additional spectral components in a sample of soft X-ray afterglow spectra. Their extra component was required to model an excess at higher energies, though, and there were no obvious humps in the light-curves, although there is some possible curvature in GRB~061110A (see http://www.swift.ac.uk/xrt$\_$curves/00238109; Evans et al. 2007, 2009).

The excess absorbing column in GRB~080307 is not strongly significant, however, and no obvious explanation presents itself for multiple components. Thus, we simply mention this other unusual feature in passing and accept that there are residual uncertainties in the measurement of excess absorption.

\subsection{Light-curve}

The X-ray light-curve of GRB~080307 showed an unusal hump starting around 100~s after the burst. Below a number of possible explanations are considered.

\subsubsection{Flare}

The XRT light-curve can be modelled as a single, unbroken power-law decay (with a large, superimposed hump), with no evidence for the series of breaks often found in the light-curves of X-ray afterglows (e.g. Nousek et al. 2006; Zhang et al. 2006). It is relatively unusual for a promptly-observed X-ray afterglow to follow a single decay, with only about 15\% of {\it Swift} XRT light-curves showing no breaks before 100~ks (Evans et al. 2009).

Flares in X-ray afterglows have a typical relative width of $\langle \Delta t /
t \rangle$~=~0.31~$\pm$~0.24 (Chincarini et al. 2007), where $\Delta t$ is the Full Width at Half Maximum (FWHM = 2.3548$\sigma$, where $\sigma$, the Gaussian standard deviation width, is the number used in the Chincarini paper) and $t$ is the time of the flare. The longest relative width in the Chincarini sample was $\Delta t / t$~=~1.27 for GRB~051117A (Goad et al. 2007a), whereas GRB~080307 has $\Delta t / t$~$\sim$~1.48. The rising emission at the start of the X-ray light-curve of GRB 080307 has
a much longer duration than is typical for such an early time relative to the
trigger. The other measurement of interest is  the relative flux variability, $\Delta F / F$, which compares the flux at the peak of the `flare' to that of the underlying power-law. In the case of GRB~080307, the values for $\Delta t / t$ and $\Delta F / F$ (1.48 and 6.17 respectively) lie in the patchy shells regime [see figure~1 in Ioka, Kobayashi \& Zhang 2005 or figure~16 in Chincarini et al. (2007)], though could also be caused by refreshed shocks.

\subsubsection{Reverse Shock}

Kobayashi et al. (2007) discuss how synchrotron self-inverse Comptonisation (SSC) can cause electrons in the reverse shock region to be upscattered into the X-ray regime (reverse shocks are typically expected to radiate photons in the optical or infrared bands). They show that SSC emission can lead to an X-ray flare around the deceleration time using GRB~050406 (Romano et al. 2006) as an example. In such a case there needs to be a steep transition from the reverse shock to the forward one ($\sim$~t$^{-(3p+1)/3}$ from figure 2 of Kobayashi et al.), but this decay phase could be very short, so can be accommodated by our data.

Reverse shock emission can only explain single flares and, in the thick shell case, is expected to overlap in time with the prompt emission. Figure~\ref{rw} demonstrates that the X-ray emission in question does follow on from the BAT detection without any obvious gap, and thus reverse shock SSC emission is a possible origin of the hump seen in this burst.

\subsubsection{Onset of afterglow}

Instead of considering the initial X-ray rise as being a flare, however, it could be thought of as the onset of the afterglow - something which was expected to be seen in the pre-{\it Swift} era [e.g., Sari \& Piran (1999a,b), Piro et al. (2005); Panaitescu \& Vestrand (2008) and Molinari et al. (2007) also discuss rising optical afterglows]. The profile of the emission in GRB~080307
is quite smooth, rising gradually, something which is not generally the case for flares formed by internal shocks. Kobayashi \& Zhang (2007) also discuss the onset of GRB afterglows, showing that a smooth bump can be produced by the forward shock emission. 

\begin{figure}
\begin{center}
\includegraphics[clip, angle=-90, width=8cm]{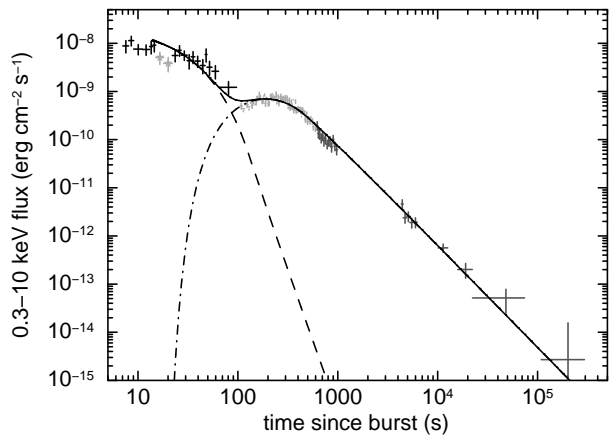}
\caption{Two exponential-to-power-law components fitted to the BAT-XRT flux light-curve. The grey stars indicate points masked out when fitting the model. Black crosses show the BAT data, while the light and dark grey points are WT and PC mode XRT data respectively. The dashed line is the fit to the prompt emission component, while the dot-dashed one models the afterglow.}
\label{rw}
\end{center}
\end{figure}

Willingale et al. (2007) discuss a method of parametrizing BAT-XRT
light-curves using one or two exponential-to-power-law functions to
model the prompt and afterglow emission; see also O'Brien et
al. (2006).  If a single function is applied to the data,
ignoring the broad hump as a flare, the start of the X-ray continuum emission is not well-fitted.  Fitting one component to
account for the prompt (BAT) emission and modelling the increase seen
in the XRT as the onset of the afterglow produces a better fit, shown in Figure~\ref{rw} [where the BAT data have been extrapolated into the 0.3--10~keV as described in O'Brien et al. (2006)]; the grey stars indicate data masked out when performing the fit. 

The power-law starts to dominate the X-ray fit $\sim$~480~s after the trigger, with a decay slope of $\alpha$~$\sim$~2.12. This is of the same order as the $\alpha$~$\sim$~1.95 found in Section~\ref{temp}.
There may be additional curvature during the later decay (after a few 10$^{4}$~s), which could be related to deceleration of a thick fireball shell (Kobayashi \& Zhang 2007 -- see their figure 7; Lazzati \& Begelman 2006). 

Very few {\it Swift}-X-ray light-curves show a rising
component (other than flares), though most show a slow-decline,
plateau phase (e.g. Nousek et al. 2006; Zhang et al. 2006). As shown
by O'Brien et al. (2006) and Willingale et al. (2007), if the
afterglow emission starts while the prompt emission is still bright,
the initial rise may be hidden; then, when the prompt component fades
sufficiently, the turn-over from the exponential to power-law decay is
seen as a `flat-ish' phase. If the XRT had started observing
GRB~080307 even 50~s later, the rising portion would have been missed,
and a (short) plateau phase would have been all that was seen at the
start of the afterglow; this was the case for GRB~080810 (Page et al. in prep). Figure~\ref{rw} shows that the function modelling the prompt, $\gamma$-ray emission drops very suddenly ($\alpha$~$\sim$~6), which may be why the afterglow was prominently visible (that is, bright with respect to the prompt emission) from an early time. The sample of bursts in Willingale et al. (2007) contains only a small number of bursts with $\alpha_{p}~>$~5. The BAT-XRT light-curve of GRB~060427 is plotted in that paper, demonstrating that, had it been a little brighter, the rise of the afterglow would probably have been briefly visible.

Following Molinari et al. (2007), the initial Lorentz factor, $\Gamma_0$, can
be estimated from the peak time of the afterglow light-curve. Here we find
that the X-ray light-curve peaks (assuming that this does correspond to the rise of the afterglow) around 234~s (Table~\ref{peak}); this corresponds to
the deceleration time. Calculating the fluence over 1--1000~keV in the rest frame to be
2~$\times$~10$^{-6}$~erg~cm$^{-2}$, we find $\Gamma$(t$_{\rm
  peak}$)~$\sim$~156($\eta_{0.2}$n$_0$)$^{-1/8}$(1+z)$^{3/8}$, where $\eta_{0.2}$ is the
radiative efficiency in units of 0.2 and n$_0$ is the particle density of the
(homogenous) surrounding medium in cm$^{-3}$. Since $\Gamma$(t$_{\rm dec}$),
which is equal to $\Gamma$(t$_{\rm peak}$) here, is
expected to be half of the initial Lorentz factor,
$\Gamma_0$~=~312($\eta_{0.2}$n$_0$)$^{-1/8}$(1+z)$^{3/8}$. 
For the mean {\it Swift} redshift of 2.26 (Jakobsson et al. 2006\footnote{The mean redshift given in the paper has been updated at http:$//$raunvis.hi.is$/$$\sim$pja$/$GRBsample.html}), this corresponds to a value of 486 (assuming typical values of $\eta$~=~0.2 and n~=~1~cm$^{-3}$ from Bloom, Frail \& Kulkarni 2003). The limit on the redshift from the Grupe et al. N$_{\rm H}$ analysis suggests an upper limit of $\Gamma_0$~$<$~562.

{\it Swift}-XRT light-curves do not typically contain smoothly rising features, although flares are seen in about 50 per~cent of bursts (e.g., Chincarini et al. 2007; Falcone et al. 2007). Utilising the light-curve repository (Evans et al. 2007, 2009), we find there are a small number of bursts where a possible early rise is seen which does not look like a standard flare (e.g. GRB~070328 -- Markwardt et al. 2007; GRB~070714B -- Racusin, Barbier \& Landsman 2007; GRB~080205 -- Markwardt et al. 2008; GRB~080229A -- Cannizzo et al. 2008;  GRB~080721 -- Marshall et al. 2008). However, in these cases the light-curve can be sufficiently well-modelled with a series of abruptly broken power-laws, rather than requiring a smooth turnover. There are also bursts which show a more gradual roll-over than can be well-modelled by a series of broken power-laws (GRB~070508 -- Grupe et al. 2007b; GRB 080319C -- Starling et al. in prep; GRB~080503 -- Mao et al. 2008; GRB~080523 -- Stroh et al. 2008;  GRB~080605 -- Sbarufatti et al. 2008). More recently, GRB~081028 (Guidorzi, Margutti \& Mao 2008; Schady \& Guidorzi 2008) showed smoothly rising emission, but in this case beginning around 10~ks after the trigger time, and reaching a peak about 10~ks later. For this burst, no spectral softening is apparent (see http://www.swift.ac.uk/xrt$\_$curves/00332851).
It remains unclear whether these bursts are displaying the same phenomenon as GRB~080307 or are unrelated, however.

\section{Summary}
\label{sum}

GRB~080307 showed a
long, smooth hump early in its X-ray afterglow light-curve. Although
superficially similar to the supernova shock breakout of GRB~060218, in this case no evidence for
radial expansion was seen. The hump is unique in the {\it Swift} dataset,
although may possibly be less well observed in a few other bursts. We have
considered various origins for the hump: a flare due to a patchy shell or
a refreshed shock, reverse shock SSC, and the onset of the afterglow. 

The long duration of the hump would make this a very slow flare for early times and the spectral softening seen at the start of the curvature is also unusual for a flare. In the case of the reverse shock interpretation, the transition to the forward shock would have to be very short to be hidden within the data.
All of these mechanisms are viable explanations of the observed
properties, however the onset of the X-ray afterglow remains a
natural interpretation.

\section{ACKNOWLEDGMENTS}
\label{ack}

The authors acknowledge support for this work at the University of Leicester by STFC. We thank Gordon Stewart for advice regarding the sky density of faint X-ray sources and the referee for helpful comments which improved the paper.

\end{document}